\documentclass[sigconf]{acmart}
\usepackage{hyperref}
\usepackage{array}
\usepackage{graphicx}
\usepackage{multicol}
\usepackage{multirow}
\usepackage{booktabs}
\usepackage[font=small,labelfont=bf]{caption}

\newcolumntype{S}{ >{\centering\arraybackslash} p{0.3in} }%
\newcolumntype{C}{ >{\centering\arraybackslash} p{0.5in} }%
\newcolumntype{P}{ p{1.0in} }%
\newcolumntype{Q}{ p{2.3in} }%

\usepackage{xcolor}

\usepackage[frozencache]{minted}
\newcommand{\cpp}[1]{{\mintinline{C++}{#1}}}

\makeatletter
\let\@float@c@listing\@caption
\makeatother

\newif\iflong
\AtBeginDocument{%
  \providecommand\BibTeX{{%
    \normalfont B\kern-0.5em{\scshape i\kern-0.25em b}\kern-0.8em\TeX}}}

\copyrightyear{2023}
\acmYear{2023}
\setcopyright{rightsretained}
\acmConference[SC-W 2023]{Workshops of The International Conference on High Performance Computing, Network, Storage, and Analysis}{November 12--17, 2023}{Denver, CO, USA}
\acmBooktitle{Workshops of The International Conference on High Performance Computing, Network, Storage, and Analysis (SC-W 2023), November 12--17, 2023, Denver, CO, USA}
\acmDOI{10.1145/3624062.3624161}
\acmISBN{979-8-4007-0785-8/23/11}

\begin{document}

%\title{Data Model Extensions for In Situ on Systems with Heterogeneous Accelerators}
\title{Extensions to the SENSEI \textit{In situ} Framework \\ for Heterogeneous Architectures}

\author{Burlen Loring}
\email{bloring@lbl.gov}
\orcid{https://orcid.org/0000-0002-4678-8142}
\affiliation{%
  \institution{Lawrence Berkeley Lab}
  \streetaddress{1 Cyclotron Rd}
  \city{Berkeley}
  \state{CA}
  \country{USA}
  \postcode{94720}
}

\author{Gunther H. Weber}
\email{ghweber@lbl.gov}
\orcid{https://orcid.org/0000-0002-1794-1398}
\affiliation{%
  \institution{Lawrence Berkeley Lab}
  \streetaddress{1 Cyclotron Rd}
  \city{Berkeley}
  \state{CA}
  \country{USA}
  \postcode{94720}
}

\author{E. Wes Bethel}
\email{ewbethel@sfsu.edu}
\orcid{https://orcid.org/0000-0003-0790-7716}
\affiliation{%
  \institution{San Francisco State University}
  %\institution{SF State University}
  \streetaddress{1 Cyclotron Rd}
  \city{San Francisco}
  \state{CA}
  \country{USA}
  \postcode{94720}
}

\author{Michael W. Mahoney}
\email{mmahoney@stat.berkeley.edu}
%\email{mmahoney@lbl.gov}
\orcid{https://orcid.org/0000-0001-7920-4652}
\affiliation{%
  \institution{Lawrence Berkeley Lab, 
  %\\ International Computer Science Institute 
  %\\ University of California at Berkeley
  %\\ \& University of California
  \\ ICSI, and UC Berkeley
  }
  %\institution{Lawrence Berkeley Lab}
  %\streetaddress{1 Cyclotron Rd}
  \city{Berkeley}
  \state{CA}
  \country{USA}
  %\postcode{94720}
}

\renewcommand{\shortauthors}{Loring, et al.}

\begin{abstract}
The proliferation of GPUs and accelerators in recent supercomputing systems, so called heterogeneous architectures, has led to increased complexity in execution environments and programming models as well as to deeper memory hierarchies on these systems.
%In addition to managing concurrent code execution among multiple on node CPUs and accelerators, effective use of these systems requires managing and marshalling data between codes running on multiple devices.
%Combined with a proliferation of PMs, some of which provide low level bare metal access to hardware, others of which provide platform portability, this leads to some interesting challenges when coupling codes developed independently from each other.
In this work, we discuss challenges that arise in \textit{in situ} code coupling on these heterogeneous architectures.
In particular, we present data and execution model extensions to the SENSEI \textit{in situ} framework that are targeted at the effective use of systems with heterogeneous architectures.
%The core data model extensions have been packaged into a new library called HAMR which implements a data model for efficient, correct, platform portable, PM agnostic, APIs for code coupling on heterogeneous architectures.
We then use these new data and execution model extensions to investigate several \textit{in situ} placement and execution configurations and to analyze the impact these choices have on overall performance.
\end{abstract}





\received{04 August 2023}
\received[revised]{XX}
\received[accepted]{XX}

\maketitle

%%%%%%%%%%%%%%%%%%%%%%%%%%%%%%%%%%%%%%%%%%%%%%%%%%%%%%%%%%%%%
\begin{listing}
\begin{minted}[fontsize=\footnotesize, linenos]{C++}
// allocate device memory
omp_set_default_device(devId);
auto devPtr = (double*)omp_target_alloc(nElem*sizeof(double), devId);

// wrap it in a shared pointer so it is eventually deallocated
std::shared_ptr<double> spDev(devPtr,
  [devId](double *ptr){ omp_target_free(ptr, devId); });

// initialize the array on the device
#pragma omp target teams distribute parallel for is_device_ptr(devPtr)
for (size_t i = 0; i < nElem; ++i)
  devPtr[i] = -3.14;

// zero-copy construct with coordinated life cycle management
auto simData = svtkHAMRDoubleArray::New("simData", spDev, nElem, 1,
                               svtkAllocator::openmp, svtkStream(),
                               svtkStreamMode::async, devId);

// pass the array to SENSEI for processing

// free up the container
simData->Delete();
\end{minted}
\caption{Packaging device data for zero-copy data transfer.}
\label{lst:zc_auto}
\end{listing}
%%%%%%%%%%%%%%%%%%%%%%%%%%%%%%%%%%%%%%%%%%%%%%%%%%%%%%%%%%%%%
%%%%%%%%%%%%%%%%%%%%%%%%%%%%%%%%%%%%%%%%%%%%%%%%%%%%%%%%%%%%%
\begin{listing}
\begin{minted}[fontsize=\footnotesize,linenos]{C++}
// this data is located in host memory, initialized to 1
auto a1 = svtkHAMRDoubleArray::New("a1", nElem, 1,
                           svtkAllocator::malloc, svtkStream(),
                           svtkStreamMode::async, 1.0);

// this data is located in device 1 memory, unitialized
omp_set_default_device(dev1);
auto a2 = svtkHAMRDoubleArray::New("a2", nElem, 1,
                           svtkAllocator::openmp, svtkStream(),
                           svtkStreamMode::async, 2.0);

// pass data to libA for the calculations on device 2
auto a3 = libA::Add(dev2, a1, a2);

// pass data to libB for I/O
auto ofile = std::ofstream("data.txt");
libB::Write(ofile, a1);
ofile.close();
\end{minted}
\caption{Illustration of PM interoperability.
%The code calls libraries \textit{libA} and \textit{libB}.
\iflong
that allocates two arrays, one on the host, the other on a device using OpenMP. This data is passed to \textit{libA} for processing. The results returned are passed to \textit{libB} for I/O.
Internally the libraries can make use of different PMs and run on different devices and none of this needs to be a part of the API or known to use the libraries.
\fi
}
\label{lst:pm_interop}
\end{listing}

%%%%%%%%%%%%%%%%%%%%%%%%%%%%%%%%%%%%%%%%%%%%%%%%%%%%%%%%%%%%%
\begin{listing}
\begin{minted}[fontsize=\footnotesize, linenos]{C++}
svtkHAMRDoubleArray*
Add(int dev, svtkHAMRDoubleArray *a1, svtkHAMRDoubleArray *a2)
{
  // use this stream for the calculation
  cudaStream_t strm = svtkStream();

  // get a view of the incoming data on the device we will use
  cudaSetDevice(dev);

  auto spA1 = a1->GetCUDAAccessible();
  auto pA1 = spA1.get();

  auto spA2 = a2->GetCUDAAccessible();                                                                                                  
  auto pA2 = spA2.get();

  // allocate space for the result
  size_t nElem = a1->GetNumberOfTuples();

  auto a3 = svtkHAMRDoubleArray::New("sum", nElem, 1,
                             svtkAllocator::cuda_async, strm,
                             svtkStreamMode::async);

  // direct access to the result since we know it is in place
  auto pA3 = a3->GetData();

  // make sure the data in flight, if it was moved, has arrived
  a1->Synchronize();
  a2->Synchronize();

  // do the calculation
  int threads = 128;
  int blocks = nElem / threads + ( nElem % threads ? 1 : 0 );
  add<<<blocks,threads,0,strm>>>(pA3, pA1, pA2, nElem);

  return a3;
}
\end{minted}
\caption{A library function in \textit{libA} that adds two arrays using the CUDA PM. 
}
\label{lst:libA}
\end{listing}
%%%%%%%%%%%%%%%%%%%%%%%%%%%%%%%%%%%%%%%%%%%%%%%%%%%%%%%%%%%%%
%%%%%%%%%%%%%%%%%%%%%%%%%%%%%%%%%%%%%%%%%%%%%%%%%%%%%%%%%%%%%
\begin{listing}
\begin{minted}[fontsize=\footnotesize,linenos]{C++}
void Write(std::ofstream &ofs, svtkHAMRDoubleArray *a)
{
  // get a view of the data on the host
  auto spA = a->GetHostAccessible();
  auto pA = spA.get();

  // make sure the data if moved has arrived
  a->Synchronize();

  // send the data to the file
  size_t nElem = a->GetNumberOfTuples();
  for (size_t i = 0; i < nElem; ++i)
    ofs << pA[i] << " ";
}
\end{minted}
\caption{A library function in \textit{libB} that writes an array to disk.
}
\label{lst:libB}
\end{listing}
%%%%%%%%%%%%%%%%%%%%%%%%%%%%%%%%%%%%%%%%%%%%%%%%%%%%%%%%%%%%%

\section{Introduction}

The current generation of heterogeneous high performance supercomputing systems provides massive computing power through a mix of CPUs, GPUs, and other accelerators such as FPGAs and AI/ML specific devices.
The use of such specialized accelerators is only expected to increase in the future~\cite{doe_het}.
Heterogeneity creates several challenges for programming and interoperability on these systems.
This includes dealing with multiple execution environments running asynchronously with respect to each other, as well as managing memory and data movement on and between accelerators and the host.
Much recent work has been devoted to platform portability, the means by which one may compile and run a single body of source code on multiple systems.
This has led to a proliferation of programming models (PMs) from which to choose, including those provided by vendors, through compilers, language standards, as well as third party solutions.
Examples of such options include OpenMP, Kokkos, Raja, HIP, SYCL, OpenCL, DPC++, std::par, and CUDA, among others.
These efforts have made it possible for code teams to cope with the rapidly evolving hardware landscape.
However, the proliferation of PMs has created additional interoperability challenges when coupling codes written by different teams.
In particular, different code teams may, for any of a number of valid reasons, choose different PMs, and they may target execution on different devices.

In terms of the data management challenges that arise when coupling independently developed codes, the challenges we increasingly face include
efficiently and correctly sending data between codes that are potentially written in different PMs and/or 
potentially processing the data on a different device or the host.
Solutions that make it possible for the codes being coupled to automatically interoperate without the need to know each others' internal capabilities and implementation details, e.g., without the need to know which PM is used and which hardware is targeted, are desirable.

Nowhere are these challenges more prevalent than in the SENSEI generic \textit{in situ} data analysis and visualization framework~\cite{sensei_api,sensei_sc16, sensei_intransit,sensei_hdf5,sensei_python}. 
SENSEI is a system that couples simulation codes to multiple back-end data processing, data transport, I/O libraries, and visualization tools through a single instrumentation; and it allows for run time switching between these back-ends.
Due to the broad diversity of simulation and back-end codes supported by SENSEI, there is a need to mediate data exchanges between simulations, potentially written in one PM and executing on one accelerator, and back-end processing codes, potentially written in a different PM and potentially executing on a different accelerator, or the host.
Because simulations are resource hungry codes, often making full use of the available memory and compute capacity, \textit{in situ} solutions such as SENSEI must be as efficient as possible. 
This means data transfers between the simulation and back end data consumer are ideally made in place, or zero-copy, whenever they can be, in order to avoid the increased memory footprint and data movement overheads associated with making a deep copy.

%%%%%%%%%%%%%%%%%%%%%%%%%%%%%%%%%%%%%%%%%%%%%%%%%%%%%%%%%%%%%%%%%%%%%%%
\begin{figure*}
    \centering
    \includegraphics[width=0.29\textwidth]{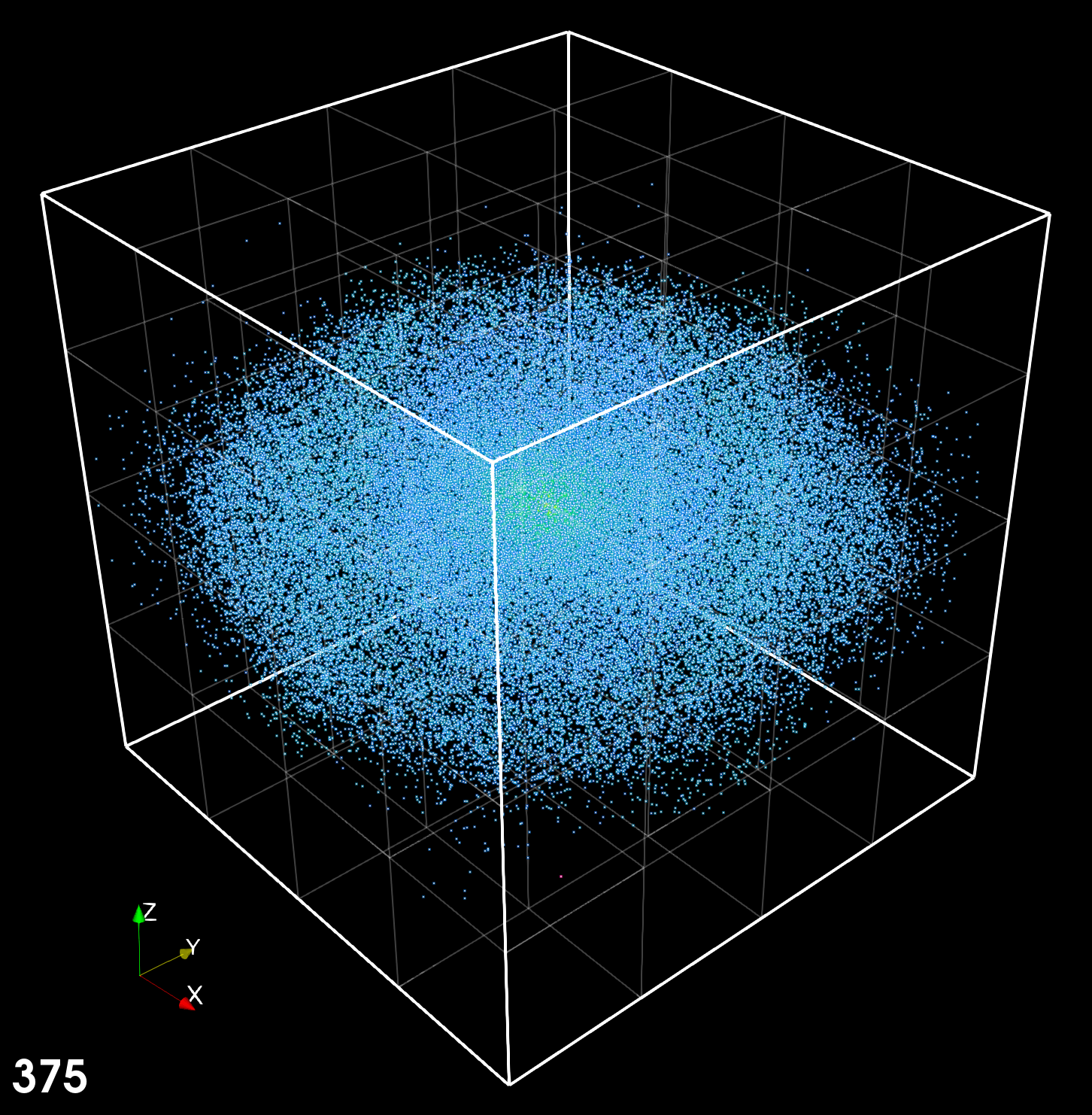}\hspace{1em}
    \includegraphics[width=0.29\textwidth]{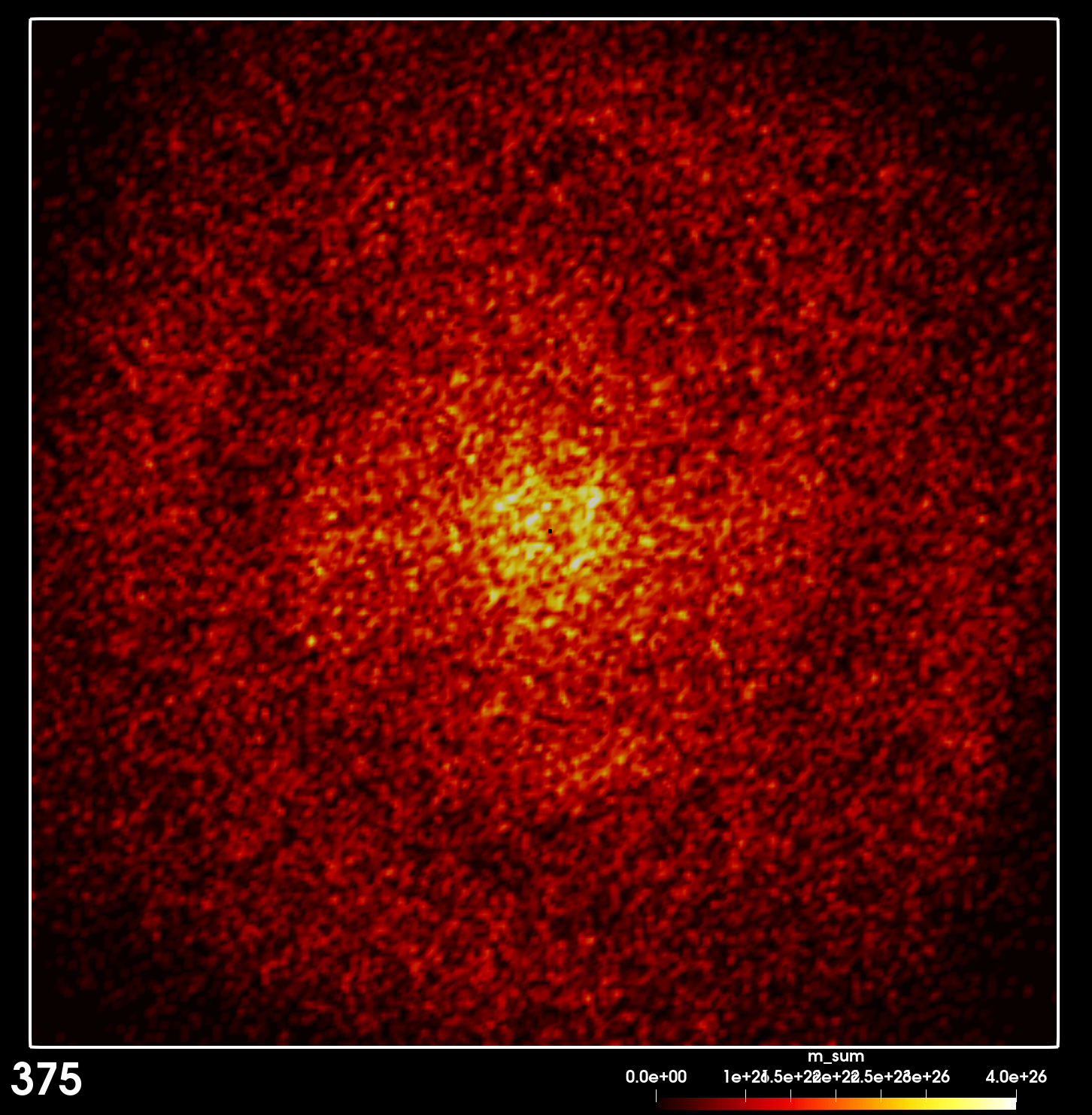}\hspace{1em}%
    \includegraphics[width=0.29\textwidth]{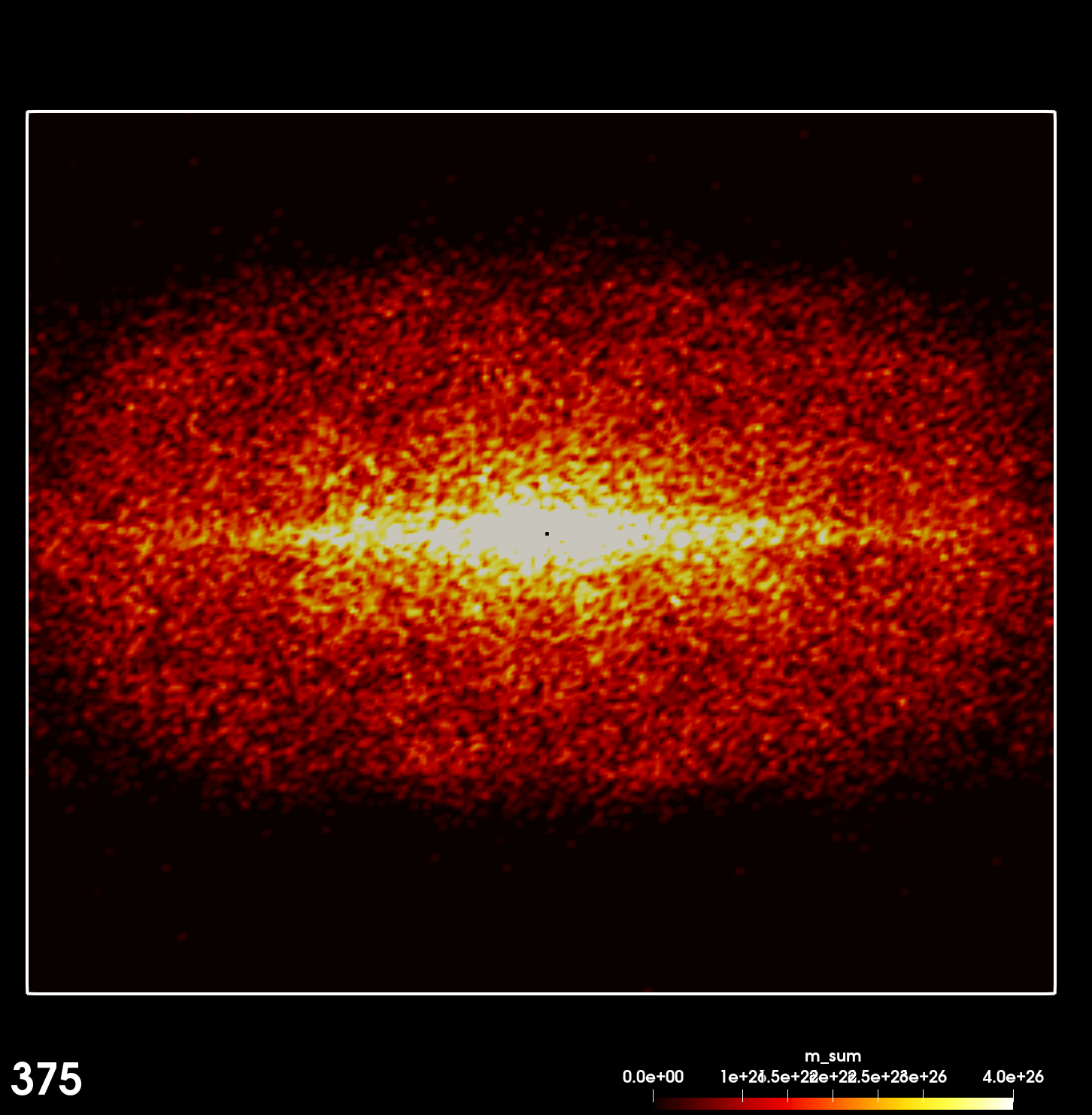}
    \caption{Left: N-body simulation of 100k bodies initialized from uniform random distributions in position, mass, and velocity with a massive body at the origin, run on 64 GPUs at step 187500. 
    %Color shows velocity magnitude. This simulation ran for 300k time steps with output and \textit{in situ} every 500 steps.
    The runs reported in Section \ref{sec:experiments} used the same initialization method with 24M bodies and ran on 512 GPUs.
    Middle: \textit{In situ} data binning in the $x$-$y$ plane of the sum of mass on a $256\times256$ grid of the same time step. Right: \textit{In situ} data binning in the $x$-$z$  plane of the sum of mass on a $256\times256$ grid of the same time step.}
    \label{fig:newton}
\end{figure*}
%%%%%%%%%%%%%%%%%%%%%%%%%%%%%%%%%%%%%%%%%%%%%%%%%%%%%%%%%%%%%%%%%%%%%%%

In this work, we present extensions to SENSEI's data and execution model to solve the challenges of \textit{in situ} code coupling on heterogeneous systems.
Section \ref{sec:sensei_dme} describes the data model extensions that address these issues and make possible correct, efficient,  data transfers through a single API.
Our work is based on new high-level data structures that provide PM interoperability as well as multi-device memory management. 
Our solution makes use of modern C++ features such as templates for efficiency and smart pointers for automated resource management.

Section \ref{sec:sensei_eme} describes the execution model extensions that enable run time scheduling of \textit{in situ} processing on different devices or the host as well as control over synchronous or asynchronous execution.
Section \ref{sec:methods} investigates different scheduling, execution, and placement options for the \textit{in situ} code, made possible by our extensions,
in the context of a simulation programmed with OpenMP target offload and an \textit{in situ} analysis programmed in CUDA.
The investigation is designed to answer the question: given a fixed number of compute nodes, each with multiple accelerators and CPU cores, what is the most effective way to utilize the available resources for \textit{in situ} processing? 
For instance, if a simulation makes heavy use of the GPUs, while CPU cores are under utilized, could overall run time be reduced by moving \textit{in situ} processing to the host? Similarly, when both simulation and \textit{in situ} make heavy use of the GPU could overall run time be reduced by moving \textit{in situ} processing to a dedicated GPU?
In Section \ref{sec:results} we analyze the impacts on the overall time to solution as well as simulation solver update times and \textit{in situ} processing times.
In Section \ref{sec:conc} we wrap up and discuss potential future directions.

\textbf{Background and Related Work.}
%
% revisions to address review comments:
While many \textit{in situ} frameworks already support the use of accelerators and GPUs (including SENSEI, ParaView Catalyst~\cite{catalyst}, VisIt Libsim~\cite{libsim}, and Ascent~\cite{ascent}), limitations in the data models of these tools preclude or limit the possibility for zero-copy data transfers of data located on the accelerator between the simulation and \textit{in situ} library to very specific scenarios.
Recent work on the Conduit \textit{in situ} data model, which is used by Catalyst and Ascent, reported zero-copy data transfer capabilities for complex mesh based data structures, with no mention or explanation of whether or how
%the data model handles
multi-accelerator heterogeneous systems are supported~\cite{conduit}.
While existing data models could potentially directly access accelerator memory through technologies like CUDA's unified memory, not all simulation codes use these mechanisms, and not all accelerator hardware supports them.
Furthermore, explicit synchronization would be required to ensure correct behavior.
None of the \textit{in situ} data models we examined handle PM interoperability or synchronization issues directly; and some rely on PM specific functionality that is not uniformly available across all PMs.

In addition to the data and execution model extensions we made to SENSEI to support heterogeneous systems, our work investigates a number of execution configurations, some of which require on node inter-accelerator or accelerator-host data movement. 
A number of previous works have compared different strategies that moved data off node through inter-process communication for processing~\cite{kress_configs, initu_sched, intransit}. 
The data binning analysis technique we used to explore our data and execution model extensions has previously been successfully used for \textit{in situ} data reduction~\cite{warpiv, kress_databin}.

VTKm is a platform portable visualization library that can run on accelerators~\cite{vtkm}.
While VTKm is widely used internally for its accelerated visualization capabilities, none of the \textit{in situ} libraries we examined use the VTKm data model in their simulation facing instrumentation APIs.

\section{Data Model Extensions}
\label{sec:sensei_dme}

In the SENSEI data model, which is based on VTK~\cite{vtkBook}, the \mintinline[breaklines,breakafter=-]{C++}{svtkD-ataArray}, an abstract base class, defines the interfaces for managing and accessing array based data.
The data sets representing mesh geometry, associated node and cell centered data, as well as the associated uncentered data are built on top of and make use of \cpp{svtkDataArray}.
However, the subclasses implementing the \cpp{svtkDataArray} APIs available in VTK are designed and implemented for host only memory management.
In order to support heterogeneous architectures, we add a new \cpp{svtkDataArray} subclass called the \cpp{svtkHAMRDataArray} (\cpp{HDA}) to the SENSEI data model.
The \cpp{HDA} provides both host and device memory management as well as PM interoperability.
Internally, \cpp{HDA} makes use of the HAMR memory management library~\cite{hamr}. 
Below, we present some of the relevant APIs along with illustrative examples.

\textbf{Initialization.} 
Before a \cpp{HDA} instance can be used it must be initialized for a particular PM and allocation strategy. 
This typically is part of the construction process, but APIs exist to initialize a default constructed instance as well.
During initialization a passed \cpp{svtkAllocator} enumeration value specifies which PM, and which specific method within the PM, is used to allocate and subsequently manage the memory.
SENSEI currently supports OpenMP offload, CUDA, and HIP allocators as well as host only allocators using \cpp{malloc}, and \cpp{new}. 
The CUDA and HIP allocators come in synchronous and asynchronous variants, variants that allocate universally addressable memory, as well as variants for allocating page locked memory.
When using asynchronous allocators, a \cpp{svtkStream} and \cpp{svtkStreamMode} must also be specified. 
\cpp{svtkStream} is a class that abstracts the differences between PM streams. 
It has automatic conversions to and from PM native streams such that these can be used interchangeably.
The \cpp{svtkStream} is used for ordering operations and explicit synchronization.
A \cpp{svtkStreamMode} enumeration value specifies a synchronization mode.
In asynchronous mode, calls to the \cpp{HDA} API return immediately while the operation is in progress making it possible to overlap allocation, data movement, and computation.
The user must add synchronization points as needed.
In synchronous mode, all operations complete before the \cpp{HDA} API call returns.
Memory is allocated on the currently active device. This provides a way to control on which device the data is located.
Line 15 of Listing \ref{lst:zc_auto} shows an example of the initialization of a newly created \cpp{HDA} instance. 

\textbf{Platform Portability and Code Execution.}
SENSEI currently supports the CUDA, OpenMP offload, and HIP PMs. 
Our data model extensions are platform portable when a platform portable programming mode is used.
The selection of a PM is left entirely to users.
Some users will opt for a platform portable option and others will opt for a vendor specific option.
Our strategy is to manage data using the selected PM and provide interoperability with all of the supported PMs so that data can be passed between any two codes, including those written in different PMs, and those targeting execution on different accelerators or the host.
Listing \ref{lst:zc_auto} illustrates platform portability achieved through the use of OpenMP. 
An example illustrating PM interoperability is given below.

\textbf{Zero-copy Data Transfer.} 
In SENSEI the simulation should always prefer a zero-copy transfer.
With zero-copy transfer, the simulation shares a pointer that gives the \textit{in situ} code direct access to the data.
The back-end data consumer can then decide
if an explicit deep copy is needed or not.
When the back-end can access the data in place, no additional work is done, reducing memory footprint and computational overhead.

Zero-copy transfer was easy to achieve on CPU-only architectures, as only a pointer and length were needed.
On heterogeneous architectures, additional information is needed, e.g., the location of the data, PM used to manage it, and possibly associated PM specific context and synchronization data structures.
\cpp{HDA}'s zero-copy data transfer APIs take pointers to externally allocated host or device memory and capture the necessary additional information.
This includes: a host or device pointer to the memory; the length of the array; an \cpp{svtkAllocator} enumeration value identifying the PM specific allocator; an integer that identifies on which device the memory currently resides; as well as a \cpp{svtkStream} and \cpp{svtkStreamMode} that are used to control ordering and synchronization of subsequent operations.
Listing \ref{lst:zc_auto} shows how data allocated and initialized using OpenMP is zero-copy transferred into a \cpp{HDA} instance.
In this example, a smart pointer is used to coordinate memory life cycle between the simulation and \cpp{HDA} instance.
We also implemented APIs that can take raw pointers. 
In that case, it's up to the user to manage memory life cycle.

%%%%%%%%%%%%%%%%%%%%%%%%%%%%%%%%%%%%%%%%%%%%%%%%%%%%%%%%%%%%%%%%%%%%%%%
\begin{table}
\centering
\noindent \begin{tabular}{ |S|C|C|S|P| }  \hline
 {\bf Num. Nodes} & {\bf \textit{In-Situ} Method}  & {\bf Ranks per node} & {\bf Ranks Total} & {\bf \textit{In-Situ}\hspace{10em}Location} \\ \toprule
 \multirow{8}{*}{128} & \multirow{4}{*}{lock step} & \multirow{2}{*}{4} & \multirow{2}{*}{512}  & all on host \\ \cline{5-5}
 & & & & on same device \\ \cline{3-5}
 & & 3 & 384 & 1 dedicated device \\  \cline{3-5}
 & & 2 & 256 & 2 dedicated devices \\ \cmidrule[1pt]{2-5}
 & \multirow{4}{*}{asynchr.} &  \multirow{2}{*}{4} & \multirow{2}{*}{512} & all on host \\ \cline{5-5}
 & & & & on same device \\ \cline{3-5}
 & & 3 & 384 & 1 dedicated device \\  \cline{3-5}
 & & 2 & 256 & 2 dedicated devices \\ \bottomrule
\end{tabular}
\caption{A summary of the runs made to investigate \textit{in situ} placement. See Section \ref{sec:results} for details.}
\label{tab:runs}
\vspace{-4pt}
\end{table}
%%%%%%%%%%%%%%%%%%%%%%%%%%%%%%%%%%%%%%%%%%%%%%%%%%%%%%%%%%%%%%%%%%%%%%%

\textbf{PM Interoperability and Location Agnostic Access.} 
We implemented an API that provides location and PM agnostic read only data access.
The purpose of the API is to make it possible to efficiently and safely pass data in between independently developed codes
which potentially make use of different PMs and potentially process data on different devices.
The code accessing the data specifies the location, on host or device, and if on device on which device, and in which PM the data will be accessed. 
If the data to be accessed is already accessible on the requested device in the requested PM, no additional work is done, and direct access is granted.
However, if the managed data is not accessible on the requested device a temporary is allocated and the data is moved. 
Any additional work required for interoperability between PMs is handled here as well.
A \cpp{std::shared_ptr} is returned from the access API so that if a temporary were used it will automatically be cleaned up when the \cpp{std::shared_ptr} goes out of scope.

Listing \ref{lst:pm_interop} illustrates PM and location agnostic data access. 
Two \cpp{HDA} instances are allocated and initialized, one on the host in line 2, and the other on device 1 using OpenMP offload in line 8.
On line 13, arrays are then passed into library \textit{libA}, which will perform an element wise addition on device 2 and return the result. 
\textit{libA} is implemented with the CUDA PM and is shown in Listing \ref{lst:libA}.
In \textit{libA}, device 2 is made active on line 8 then lines 10 and 13 use the \cpp{HDA} access API to obtain a view of the data to operate in the CUDA PM.
If any host-device or inter-device data movement, or PM interoperability transformations, are needed, these are handled automatically in the \cpp{HDA} access API invisibly to \textit{libA}.
In this way \textit{libA} can operate on data in any location managed by any PM with the same code.
On line 24 \textit{libA} uses the more efficient direct access API to get raw pointer to memory for the result because the location and PM are known.

Back in Listing \ref{lst:pm_interop}, the result of the calculation made in \textit{libA} is returned on line 13. A file is created and \textit{libA}'s result is passed to library \textit{libB} for output to disk. 
The implementation of \textit{libB} in host only C++ is shown in Listing \ref{lst:libB}.
\textit{libB} calls the \cpp{HDA} access API on line 4 to get a view of the data to operate on the host. 
Any host-device data movement is handled automatically and invisibly to \textit{libB} if it is needed.
This example demonstrates how our data model handles PM interoperability and data movement automatically.

\section{Execution Model Extensions}
\label{sec:sensei_eme}
In this section we describe two additions to SENSEI's execution model we made in order to support execution on heterogeneous architectures. 
The first is for the specification of an execution method. 
The second implements a number of \textit{placement} options that give run time control over on which accelerator or the host the \textit{in situ} code runs.
These new features are exposed to users through an API and SENSEI's run time XML configuration.

The new execution methods are: \textit{lockstep} where the simulation and \textit{in situ} code take turns; and \textit{asynchronous} where the \textit{in situ} code uses threading to execute concurrently with the simulation.
In terms of control over \textit{in situ} placement, we implemented means for both manual explicit device selection and automatic device selection.
Automatic device selection uses a number of run time provided control parameters along with the process's MPI rank and the number of on node devices to select a device to execute on according to the following rule:
\vspace{-2pt}
\begin{equation} \label{eqn:auto_dev}
d = ( r \bmod n_u * s + d_0 ) \bmod n_a  ,
\end{equation}
where: $d$ is the assigned device; $r$ is the MPI rank of the process making the query; $n_u$ is the number of devices to use per node; $s$ is the stride, $d_0$ is the offset, and $n_a$ is the total number of device available on the node.
$r$ and $n_a$ are initialized from system queries, while $n_u$, $s$, and $d_0$ can optionally be specified by the user. 
By default, $n_u = n_a$, $s = 1$, and $d_0 = 0$.
\iflong
The new XML attributes are described in the following table.

\noindent \begin{tabular}{|l|Q|} \hline
\textbf{attribute}  & \textbf{description} \\ \hline
device\_id  &  controls the device id to execute on. -2 for automatic assignment using equation \ref{eqn:auto_dev}. -1 for host. 0 to num devices for explicit manual selection. default -2. \\ \hline
device\_start & the offset $d_0$ in equation \ref{eqn:auto_dev}. default 0. \\ \hline
device\_stride & the stride $s$ in equation \ref{eqn:auto_dev}. default 1. \\ \hline
devices\_to\_use & the number of devices to use per node. default the number of devices per node. \\ \hline
async & when set \textit{in situ} processing should run asynchronously with respect to the simulation. \\ \hline
\end{tabular}
\fi
\noindent The new control parameters and API are defined in the base class for SENSEI analysis back-ends and therefor available to all back-ends.
%

% results fig as submitted
%%%%%%%%%%%%%%%%%%%%%%%%%%%%%%%%%%%%%%%%%%%%%%%%%%%%%%%%%%%%%%%%%%%%%%%
\begin{figure}
\centering
\includegraphics[width=0.4\textwidth]{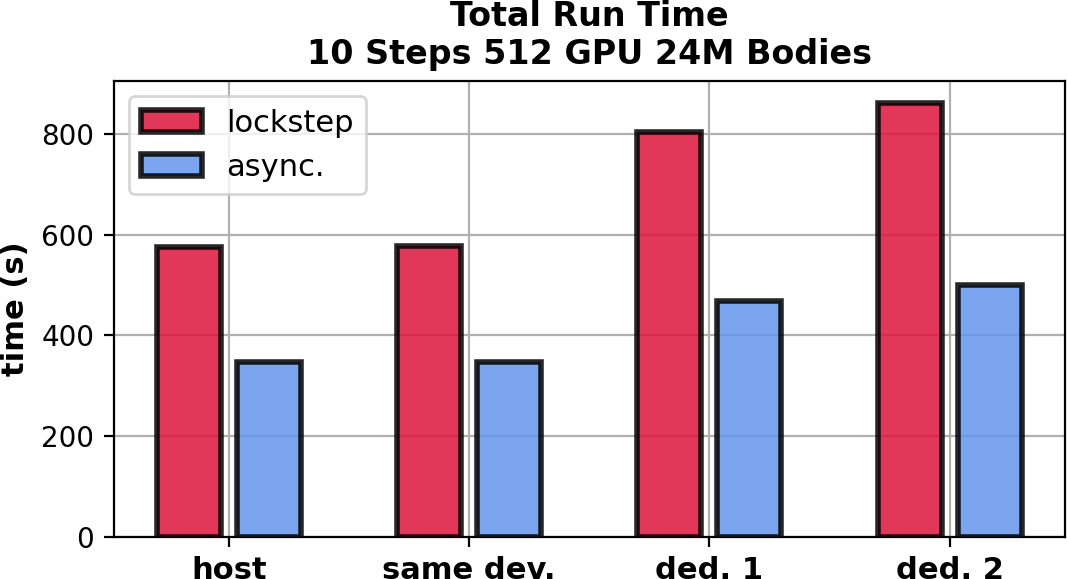}
\caption{Total run time for lockstep(red) and asynchronous(blue) \textit{in situ} for each of four \textit{in situ} placements.}
\label{fig:results_tot}
\end{figure}
%%%%%%%%%%%%%%%%%%%%%%%%%%%%%%%%%%%%%%%%%%%%%%%%%%%%%%%%%%%%%%%%%%%%%%%
%%%%%%%%%%%%%%%%%%%%%%%%%%%%%%%%%%%%%%%%%%%%%%%%%%%%%%%%%%%%%%%%%%%%%%%
\begin{figure}
\centering
\includegraphics[width=0.4\textwidth]{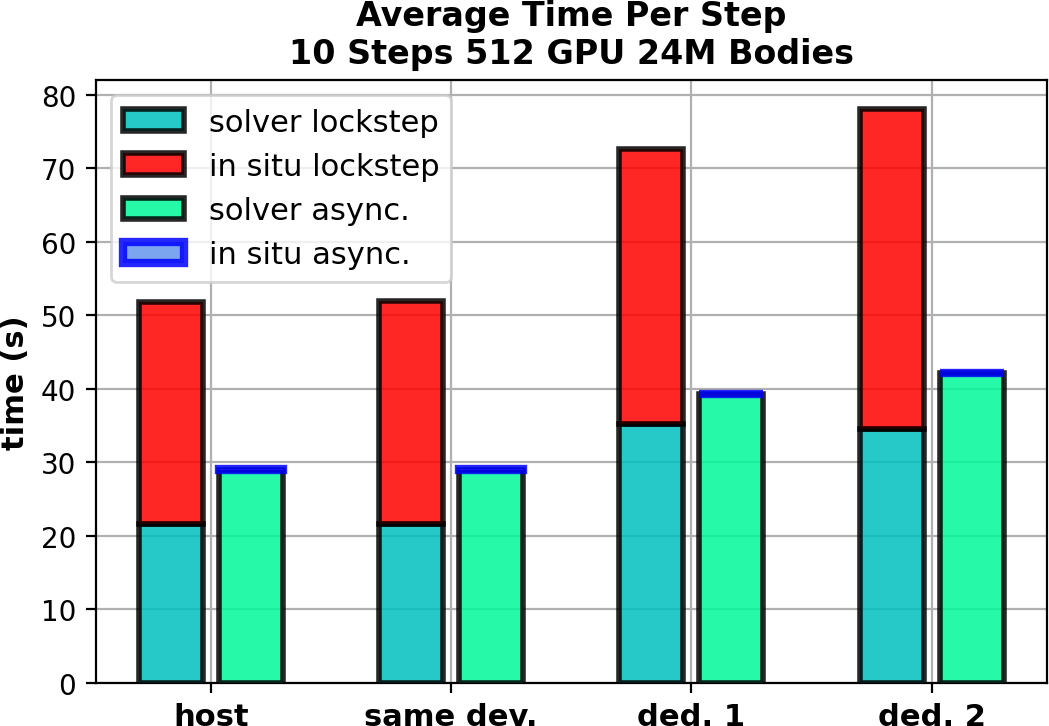}
\caption{The average time per iteration of the solver and \textit{in situ} processing, for each of the four \textit{in situ} placements. Asynchronous \textit{in situ} execution times are blue, while lockstep \textit{in situ} times are red. Solver times are cyan. }
\label{fig:results_avg}
\end{figure}
%%%%%%%%%%%%%%%%%%%%%%%%%%%%%%%%%%%%%%%%%%%%%%%%%%%%%%%%%%%%%%%%%%%%%%%

\section{Evaluation of the Extensions}
\label{sec:methods}
In this section, we report a set of empirical results in which we seek to answer the question:
``How can we best make use of multiple on node accelerators and CPUs for \textit{in situ} processing?''
We also demonstrate PM interoperability between a simulation written for OpenMP offload and an \textit{in situ} analysis code written in CUDA, coupled through the SENSEI generic \textit{in situ} framework.

\subsection{Simulation}
\label{sec:newton}
In our experiments we used the Newton++ simulation code~\cite{newtonpp}.
Newton++ is an open source direct n-body simulation with a second order, time reversible, symplectic integration scheme. 
Newton++ is written in C++ and parallelized with MPI and OpenMP device offload. Each MPI rank owns a unique spatial subdomain of the simulated volume and is responsible for integrating bodies within its subdomain.
As bodies evolve in time, a repartitioning phase migrates bodies that have moved out side of a given subdomain to the correct MPI rank.
Newton++ is instrumented with SENSEI, and it has a VTK compatible output format for post processing and visualization.
Newton++ also supports initial conditions generated by MAGI, the Many-component Galaxy Initializer~\cite{magi}, as well as initialization from uniform random distributions in position, mass, and velocity.
The leftmost panel in Figure~\ref{fig:newton} shows example of the data generated by Newton++.

\subsection{\textit{In Situ} Data Binning}
\label{sec:databin}
Given tabular data where columns represent different variables and rows represent co-occurring measurements or realizations of these variables, data binning specifies a subset of the variables to use as the coordinate axes of a uniform Cartesian mesh and transforms the data into the new coordinate system.
For each realization, the values of the coordinate variables locate the mesh cell, or bin, to which the realization belongs.
The low and high bounds of the mesh axes can be manually specified or obtained on the fly by calculating the minimum and maximum of the respective coordinate variables. 
Incrementing a per-mesh-cell counter creates a histogram of the data in terms of the chosen coordinates. Additional reduction operations are used to incorporate, or bin, non-coordinate variables into the result. The reduction operations we support are summation, minimum, maximum, and average.

For example, the first panel in Figure~\ref{fig:newton} shows data produced in a 100k body n-body simulation;
the middle panels shows the result of data binning body mass with summation onto a $256 \times 256$ mesh with $x$ and $y$ body positions used as coordinate axes; and  
the right panel shows the same with body $x$ and $z$ positions used as coordinate axes.
These examples show the use of the bodies' spatial coordinates as binning axes. However, it is common to use other per-body attributes, such as momentum or velocity, as the coordinate axes.

Our implementation is parallelized with MPI and CUDA.
We provide a CPU implementation that runs on the host as well as a CUDA implementation that runs on an assigned device.
Both implementations can run asynchronously in a C++ thread.
We make use of the SENSEI execution model extensions described in Section \ref{sec:sensei_eme} for placement and execution method control.
The extensions to the data model described in Section \ref{sec:sensei_dme} are used for data access and management. 

\subsection{Empirical Evaluation}
\label{sec:experiments}
In our empirical evaluation, we run the n-body simulation on a fixed number of GPUs on NERSC's supercomputer Perlmutter while changing how resources are allocated between simulation and \textit{in situ} processing.
We investigate four \textit{in situ} placements in conjunction with two \textit{in situ} execution methods for a total of eight cases.

The four \textit{in situ} placements are: 1. all on host; 2. on the same device; 3. on a dedicated device; 4. on two dedicated devices.
For all four \textit{in situ} placements each simulation rank is assigned a specific GPU, there is always only 1 simulation rank per GPU.
For the \textit{host} placement, \textit{in situ} calculations are scheduled on the host.
Data is moved from the GPU on which it is generated to the system's main memory banks for \textit{in situ} processing. 
With the \textit{same device} placement, \textit{in situ} calculations are scheduled on the device where it is generated.
Four MPI ranks per-node, one per GPU, are used with the \textit{host} and \textit{same device} placements.
With the \textit{dedicated device} placement, one GPU per node is reserved for \textit{in situ} processing. The three remaining GPUs per node are used exclusively by the simulation. In this placement three MPI ranks are used per node.
Data is moved from the three simulation GPUs per node to the one \textit{in situ} GPU per node for processing.
With the \textit{2 dedicated devices} placement each MPI rank has one GPU dedicated to simulation and one GPU dedicated to \textit{in situ} processing.
Data is moved from the simulation to its paired GPU for processing. In this placement two MPI ranks per node are used.

We investigated two execution methods: \textit{lockstep}, and \textit{asynchronous}.
With \textit{lockstep} execution the simulation and \textit{in situ} processing take turns with the simulation waiting for the \textit{in situ} to fully complete before proceeding.
The \textit{lockstep} method makes zero-copy copy data access possible in some of the cases.
With \textit{asynchronous} execution, the \textit{in situ} analysis code runs in a separate thread, asynchronously with respect to the simulation.
The \textit{in situ} code deep copies the relevant data, launches a thread for \textit{in situ} processing, and returns immediately to the simulation. The simulation and \textit{in situ} processing then proceed concurrently.

All runs were made on 128 Perlmutter nodes, using 512 GPUs.
The various \textit{in situ} placements allocated GPUs differently between \textit{in situ} and simulation. 
The runs exercising the 8 cases described above are summarized in table \ref{tab:runs}.
In all runs, Newton++ was configured with 24M bodies generated from the uniform random initial condition. \textit{In situ} processing via SENSEI was performed at every iteration. I/O for \textit{post hoc} visualization and body repartitioning were disabled during the runs.
During \textit{in situ} processing the data binning operator was applied to 10 variables
over 9 coordinate systems
for a total of 90 binning operations.
Binning of each coordinate systems was done sequentially in a separate data binning operator instance and orchestrated by SENSEI using its XML configuration feature. The SLURM batch scripts and XML configs used in the experiments are provided in Appendix \ref{sec:appendix}.

\subsection{Results and Discussion}
\label{sec:results}
Figures \ref{fig:results_tot} and \ref{fig:results_avg} summarize the data gathered in the eight experimental runs.
Figure \ref{fig:results_tot} shows the total run time including initialization, solver steps, \textit{in situ} processing, and finalization. 
Figure \ref{fig:results_avg} shows the average time per iteration spent in the solver and \textit{in situ} processing in a stack plot. The total height is the average time spent per iteration.
Both plots organize the data into four groups, one for each \textit{in situ} placement. Within each of the four groups, one bar represents the \textit{lockstep} execution method and the other \textit{asynchronous} execution method.

These results show that, across all placements, executing \textit{in situ} asynchronously is beneficial and reduced the total run time.
The apparent time spent in \textit{in situ} processing when \textit{asynchronous} execution was used was very small, less than $10ms$ across all time steps and all placements. 
This makes it look like \textit{in situ} is effectively free when executing asynchronously. 
However, comparing the solver time between the \textit{lockstep} and \textit{asynchronous} cases in Figure~\ref{fig:results_avg} it is clear that the solver was slowed down across all placements when the \textit{in situ} was executed asynchronously.
None the less running the \textit{in situ} asynchronously significantly reduced the total run time.

The placements assigning one or two dedicated devices for \textit{in situ} processing made use of a reduced total number of MPI ranks, as well as a reduced total number of GPUs used by the simulation and \textit{in situ} respectively.
The reduced levels of concurrency led to longer run times.
For instance the \textit{two dedicated devices} placement used only half of the MPI ranks that were used in the \textit{same device} placement, and individually both simulation and \textit{in situ} had access to half the number of GPUs.
There was a negligible difference between the \textit{host only} and \textit{same device} placements.
We observe that data binning is not an ideal algorithm for GPUs since it requires the use of atomic memory updates to deal with races between GPU threads accessing the same bin.

\section{Conclusions and Future Work}
\label{sec:conc}

We have presented data and execution model extensions to the SENSEI \textit{in situ} framework to automate data management and PM interoperability on heterogeneous architectures.
We demonstrated the PM interoperability, inter-device memory management, and on node \textit{in situ} placement capabilities provided by our extensions in a set of empirical evaluations
that investigated different \textit{in situ} placement and execution configurations
on NERSC's Cray Perlmutter.

The asynchronous execution method we implemented resulted in overall reduced run time in spite of slowing the solver down relative to  the lockstep method.
In future work we plan to do deeper profiling to understand this better as well as
more profiling to better understand the opportunities for improving performance when assigning one or two dedicated devices for \textit{in situ} processing.
We will profile and optimize the data binning implementation to achieve a speed up on the GPU relative to the CPU.
We will also add support for SYCL as well as third party PMs such as Kokkos.

\begin{acks}
%short and sweet
This material is based upon work supported by the U.S. Department of Energy, Office of Science, Office of Advanced Scientific Computing Research under Award Number DE-AC02-05CH11231. This research used resources of the National Energy Research Scientific Computing Center (NERSC), a U.S. Department of Energy Office of Science User Facility.
% Longer and better ack provided by Gunther.
%This work was supported by the U.S. Department of Energy, Office of Science, Office of Advanced Scientific Computing Research under Award Number DE-AC02-05CH11231 through the grant “Scalable AnalysisMethods and In Situ Infrastructure for Extreme Scale Knowledge Discovery,” program manager Dr. Margaret Lentz. This research used resources of the National Energy Research Scientific Computing Center (NERSC), a U.S. Department of Energy Office of Science User Facility located at Lawrence Berkeley National Laboratory, operated under Contract No. DE-AC02-05CH11231 using NERSC award ASCR-ERCAP-0023337.
\end{acks}

\bibliographystyle{ACM-Reference-Format}
\bibliography{isav_2023_insitu_scheduling}

\appendix
\section{Reproducibility}
\label{sec:appendix}
We have organized all of the information needed to reproduce our experiments in a repository on github. \href{https://github.com/SENSEI-insitu/ISAV_2023}{https://github.com/SENSEI-insitu/ISAV\_2023}
The information is organized as follows:

\noindent \begin{tabular}{|l|l|} \hline
\textbf{Directory} & \textbf{Description} \\ \hline
software & documents the versions of software used \\ \hline
environment & documents the requisite Perlmutter modules \\ \hline
build & Documents the configure and build steps \\ \hline
sensei\_xml & SENSEI XML configurations used in the runs \\ \hline
run & SLURM batch scripts used in the runs \\ \hline
data & data gathered in the runs \\ \hline
analysis & scripts used to analyze the runs \\ \hline
\end{tabular}

\end{document}